 \documentclass[final,5p,times,twocolumn]{elsarticle}

\usepackage{graphicx}
\usepackage{subcaption}
\usepackage{multirow}
\usepackage{wrapfig}
\usepackage{array}

\usepackage{tabularray}
\usepackage{tabularx}
\usepackage{rotating}
\usepackage{makecell}
\usepackage{booktabs}

\usepackage[utf8]{inputenc}
\usepackage{amsmath}
\usepackage{amsthm}
\usepackage{amsfonts}
\usepackage{epsfig}
\usepackage{psfrag}
\usepackage{pstricks}
\usepackage{algorithm}
\usepackage{stfloats}

\usepackage{cuted}

\usepackage{algorithmicx}
\usepackage{algpseudocode}  
\floatname{algorithm}{Algorithm}

\graphicspath{{./Figures/}}

\usepackage{amssymb}
\usepackage{bm}
\usepackage{hyperref}
\hypersetup{
            colorlinks=true,
            linkcolor=blue,
            anchorcolor=blue,
            citecolor=blue
            }

\biboptions{comma,sort&compress}

\usepackage{enumitem}
\setlist{nosep,topsep=-\parskip}

\usepackage{ulem}

\journal{Elsevier}

\begin{document}
\baselineskip11pt

\begin{frontmatter}

\title{SplineGen: a generative model for B-spline approximation of unorganized points}
\author[]{Qiang Zou\corref{cor}}\ead{qiangzou@cad.zju.edu.cn}
\author{Lizhen Zhu}
\cortext[cor]{Corresponding author.}
\address{State Key Laboratory of CAD\&CG, Zhejiang University, Hangzhou, 310027, China}

\begin{abstract} 
This paper presents a learning-based method to solve the traditional parameterization and knot placement problems in B-spline approximation. Different from conventional heuristic methods or recent AI-based methods, the proposed method does not assume ordered or fixed-size data points as input. There is also no need for manually setting the number of knots. It casts the parameterization and knot placement problems as a sequence-to-sequence translation problem, a generative process automatically determining the number of knots, their placement, parameter values, and their ordering. Once trained, SplineGen demonstrates a notable improvement over existing methods, with a one to two orders of magnitude increase in approximation accuracy on test data. 
\end{abstract}


\end{frontmatter}

\section{Introduction} 
\label{sec:introduction}
Generative artificial intelligent (AI) models have proven to be very powerful in natural language processing, computer vision, and computer graphics~\cite{vaswani2017attention}. In particular, there is emerging use for the synthesis of 3D shapes lately, such as meshGPT~\cite{siddiqui2023meshgpt}, SolidGen~\cite{jayaraman2022solidgen}, and ComplexGen~\cite{guo2022complexgen}. 
However, there has been limited use of these models, or even the broader deep learning models, in computer-aided design (CAD) modeling. This is mainly because those models are probabilistic (and therefore inherent uncertainty therein), while all CAD algorithms require high certainty and accuracy. 
This work aims to show that, despite this gap, generative models offer considerable potential for improving classic CAD algorithms by replacing the expertise-based, heuristic parts in classic algorithms with data-driven methods.

Approximating a set of data points with a B-spline curve~\cite{de1972calculating} is the algorithm of choice. It is a fundamental algorithm in CAD/CAM, underlying many high-level modeling operations such as boundary evaluation, Booleans, features, tool path generation, reverse engineering, shape optimization etc.~\cite{stroud2011solid,zou2023variational,liu2021memory,zou2019push,wang2023quasi,zou2021length,zou2014iso,li2023xvoxel,zou2024meta,su2020accurate}. The approximation involves three main procedures: parametrization, knot placement, and approximation error minimization~\cite{PiegTill1996}. The first two are prerequisites for running the third. 
Parametrization means associating each data point with a parameter value.
Knot placement refers to the choice of the number of knots and their locations for constructing spline bases.
It is well known that the parameterization and knot placement have a significant impact on the final approximation accuracy~\cite{scholz2021parameterization}.

Existing parameterization and knot placement methods basically fall under two different strategies. The first strategy heuristically chooses a parameterization and knot placement \textit{a posteriori} for given data points. 
Typical examples include uniform parameterization~\cite{PiegTill1996}, chord length parameterization~\cite{fang2013improved}, centripetal parameterization~\cite{lee1989choosing}, and Foley-Nielson parameterization~\cite{foley1989knot}, among many others~\cite{balta2019dynamic}.
For knot placement, uniform knots~\cite{PiegTill1996}, the knot placement technique (KTP)~\cite{PiegTill1996} and the new KTP (NKTP)~\cite{piegl2000surface} are typical examples. These methods have also been augmented by incorporating various optimization techniques, e.g., ~\cite{yong2022unimodality}.
Like all heuristic methods, those mentioned above can provide acceptable results in many but not all of the potential cases. 
The second strategy learns a neural network to directly map data points to their corresponding parameters and knots, in an end-to-end manner~\cite{scholz2021parameterization}. Fully connected neural networks (FCNN)~\cite{laube2018deep} and residual networks (ResNet)~\cite{scholz2021parameterization} have been used for this purpose so far. Reduced approximation errors were demonstrated, but the application of these methods is consistently limited to ordered and fixed-size data points.

In this work, we propose to cast the problems of parameterization and knot placement as a generative sequence-to-sequence translation problem, where unorganized input becomes possible. Rather than directly adopting existing networks (e.g., FCNN and ResNet), this work develops an extended version of the existing Transformer network~\cite{vaswani2017attention} to meet the specific needs of B-spline approximation. The new generative model, SplineGen, can automatically determine the number of knots, their placement, parameter values, and their ordering. More specifically, SplineGen has the following features:
\begin{enumerate}[label=(\arabic*)] 
    \item \textbf{Size-independent}. Restricting input data points, parameters, and knots to a fixed size would limit the practical significance of learning-based methods. The point partitioning trick as used in~\cite{scholz2021parameterization,laube2018deep} may help but cannot solve the problem altogether. By contrast, the proposed method provides a principled way to handle variable-size input data points, parameters, and knots through the use of autoregressive models.
    
    \item \textbf{Point permutation invariant}. In many practical scenarios, input points may not be in a well-ordered position as required by most existing methods. As a result, a B-spline approximation-oriented neural network should exhibit permutation invariance over the input. That is, the output parameters and knots remain unchanged when input data points are permuted. This problem is to be solved by extending the transformer network with an additional self-organization module.
    
    \item \textbf{Parameter-knot alignment.} Many existing methods consider parameterization and knot placement problems separately. However, they correlate highly with each other, and lower approximate errors will be achieved if they work in a cooperative way, refer to Section~\ref{sec:problem-statement} for more details. This necessitates a neural network that can align the generation of parameters and knots. To do so, this work uses a shared autoencoder model to learn shared point embeddings for both tasks. Furthermore, we add a new module called internal cross-attention to the parameter decoder and the knot decoder to explicitly align parameter generation and knot generation. 
    
    \item \textbf{High robustness.} For a set of input data points, removing some of them or adding more sample points from the same curve should not affect the output significantly, if not completely identical. This simply means robustness towards data point removal or addition (of course, without changing the unknown underlying curve). To achieve this, we further extend the transformer network by adding a masking mechanism for simulating point removal/addition and predicting the corresponding output masks.    
\end{enumerate}

The remaining content is organized as follows: Sec.~\ref{sec:related-work} provides a review of related research studies. Sec.~\ref{sec:problem-statement} states the problem to be solved, and Sec.~\ref{sec:methods} elaborates the proposed SplineGen network. Examples and comparisons with existing methods are provided in Sec.~\ref{sec:results}, followed by conclusions in Sec.~\ref{sec:conclusion}.

\section{Related Work}
\label{sec:related-work}
Spline techniques are fundamental to the modeling of curves, surfaces, structures, as well as semantic shapes such as tool paths, in CAD/CAM applications~\cite{zou2022robust,zhao2024tpms2step,zou2020decision,wang2023computing,zou2019variational,luo2023simple,zou2013iso,ding2021stl,zou2021robust}. For this reason, a ton of related studies have been presented in the literature. They may be divided into two categories: the traditional and the recent learning-based. Because there have already been excellent reviews on the method in the traditional category, e.g.,~\cite{scholz2021parameterization,fang2013improved}, this paper will only gives a brief summary of those methods.

\subsection{Traditional methods}
\textbf{Parametrization methods.} 
Because of the importance of parameterization in B-spline approximation, many research studies on this topic have been carried out. Recently, Scholz and J\"uttler provided an informative summary on the classification and development of those methods in their paper~\cite{scholz2021parameterization}. Basically, heuristic methods are the dominant means of parameterization, e.g., uniform~\cite{PiegTill1996}, chord length~\cite{fang2013improved}, centripetal~\cite{lee1989choosing}, universal~\cite{lim1999universal}, and Foley-Nielson~\cite{foley1989knot}, among many others. 
Such basic versions have also been used as an initial guess for the parameterization and then optimized to achieve better approximation results, e.g.,~\cite{gao2019deepspline}. Typically, this leads to a highly non-linear constrained optimization problem. One line of relevant research studies focuses on the solving algorithms for this optimization problem, e.g., particle swarm optimizers~\cite{iglesias2015bat, iglesias2016four}, and global optimizers~\cite{speer1998global}. The other line designs various error metrics to control the ``goodness" of the final parameterization, e.g., squared distance errors~\cite{wang2006fitting} and angular speed uniformness~\cite{yang2013improving}.

\textbf{Knot placement methods.} Existing knot placement methods may be classified into two types: knot assignment that computes the whole knot vectors simultaneously; and knot refinement that iteratively modifies initial knots until an error threshold is met. To reflect the distribution of data points and guarantee that every knot span includes at least one parameter, there are knot assignment methods like AVG (averaging technique) (for point number $m$ = control point number $n$) and KTP (knot placement technique) (for point number $m>$ control point number $n$)~\cite{PiegTill1996}. Because KTP will have stability issues when $m$ is slightly greater than $n$. Piegl and Tiller~\cite{piegl2000surface} proposed a stability-enhanced version (NKTP) of KTP by using an averaging method. Like in parameterization, these methods have also been used as initial guesses and then optimized by iterative knot insertion~\cite{luo2019knot, michel2021new}. The essential task here is to determine when and where to insert the new knots. Several heuristics have been developed, such as angle variations~\cite{li2005adaptive}, curvatures~\cite{park2007b}, and minimum knot size~\cite{kang2015knot}. Evolutionary optimization algorithms~\cite{ yoshimoto2003data, luo2022knot} and neural network-based optimization~\cite{wen2024deep} have also been used to determine the knot insertion position.

The methods discussed above have proven effective in many applications, but due to their heuristic nature, they cannot provide acceptable results in all of the potential cases. This motivates the recent use of machine learning methods to carry out parameterization and knot placement, in an end-to-end manner.

\subsection{Learning-based methods}
\textbf{Parametrization methods.} The objective here is to learn a neural network to directly map data points to their corresponding parameters. The method presented in~\cite{laube2018deep} appears to be the first to do so. A standard FCNN with a fixed input size of 100 data points and three hidden layers, each with 1,000 nodes, was used to construct the mapping. Because the input size is fixed, up-sampling or down-sampling is needed to run the model if the given number of data points is not equal to 100. Moreover, the training and inference of the model are time-consuming. To overcome this limitation, Scholz and J\"uttler~\cite{scholz2021parameterization} replaced FCNN with ResNet, which has a fixed input size of $2d+1$ data points if the underlying curve is of degree $d$. Multiple models need to be trained, each for a degree. Each model can only parameterize $2d+1$ data points. If there are more than $2d+1$ data points, heuristical down/up-sampling is required. By contrast, the method in this paper can directly parameterize an arbitrary number of data points. 

\textbf{Knot placement methods}. In the same paper~\cite{laube2018deep}, FCNN is also used to predict knot positions. The model has a fixed input size of 100 data points and three hidden layers, each with 500 nodes. It assumes that knots are a subset of parameters. Under this assumption, predicting knot positions is converted to an easier classification problem on parameters. A similar idea has also been used in~\cite{laube2018learnt}, where a support vector machine is used to carry out the classification instead of FCNN. It is, however, arguable to snap knots onto parameters since knots and parameters do not have a shared relationship in many cases. In addition, their method's input size is kept fixed.

\textbf{3D generative methods.} The proposed method in this work makes use of generative models, so it is closely related to the emerging field of generative 3D modeling~\cite{shi2022deep}. Generative models have been successfully applied to a series of shape representations such as voxels~\cite{schwarz2022voxgraf}, point clouds~\cite{zeng2022lion}, neural implicit fields~\cite{gao2022get3d, cheng2023sdfusion}, meshes~\cite{siddiqui2023meshgpt}, and solids~\cite{jayaraman2022solidgen,guo2022complexgen}. However, there has been limited use of generative models for B-spline modeling. Despite this, the way they designed the networks is very inspiring to this work. In particular, SplineGen has the same high-level network architecture (i.e., encoder-decoder) as most existing generative models, the low-level design of the encoder module and decoder module is different though. 

Existing learning-based parameterization and knot placement methods have proven that they can be a promising alternative to the traditional heuristic methods in B-spline approximation. And reduced approximation errors were demonstrated. Nevertheless, there is still room for improvement, especially in extending the applicability and increasing accuracy. Following what has been done in~\cite{laube2018deep,scholz2021parameterization}, this work presents a generative parameterization and knot placement, with added features of unorganized data points, size-independent, point permutation invariant, knot-parameter alignment, and high robustness.
  
\section{Problem statement}
\label{sec:problem-statement}
Mathematically, a B-spline curve is:
\begin{equation}
    C(u) = \sum_{i=0}^{n} N_{i,p}(u) P_i
\end{equation}
where $u$ is a varying parameter in $[0,1]$, $N_{i,p}(u),\ i=0,\dots,n,$ the B-spline basis functions of degree $p$, and $P_i, \ i=0,\dots,n,$ the control points. The basis functions $N_{i,p}(u)$ are generated using a knot vector $T = [t_0, \dots, t_{n+p+1}]$, as follows:
\begin{equation}
    \begin{aligned}
        &N_{i,0}(u)= 
            \begin{cases}
                1,\quad &t_i\leq{u}<t_{i+1} \\
                0,\quad &otherwise
            \end{cases} \\
        &N_{i,p}(u)=\frac{u-t_i}{t_{i+p}-t_i}N_{i,p-1}(u)+\frac{t_{i+p+1}-u}{t_{i+p+1}-t_i}N_{i+1,p-1}(u)
    \end{aligned}  
\end{equation}

Approximating data points ${D_j},\ j=0,\dots,m,$ with a B-spline curve is to solve the following optimization problem:
\begin{equation}\label{eq:optimizatioin}
    \min_{P_0,\dots,P_n} \sum_{j=0}^{m} || D_j - \sum_{i=0}^{n} N_{i,p}(u_j) P_i ||^2_2 
\end{equation}
where $u_j$ is the parameter value corresponding to the data point $D_j$. Constructing the mapping from $D_j$ to $u_j$ is the so-called parameterization. 

\begin{figure}[htbp]
    \centering
    \includegraphics[width=0.49\textwidth]{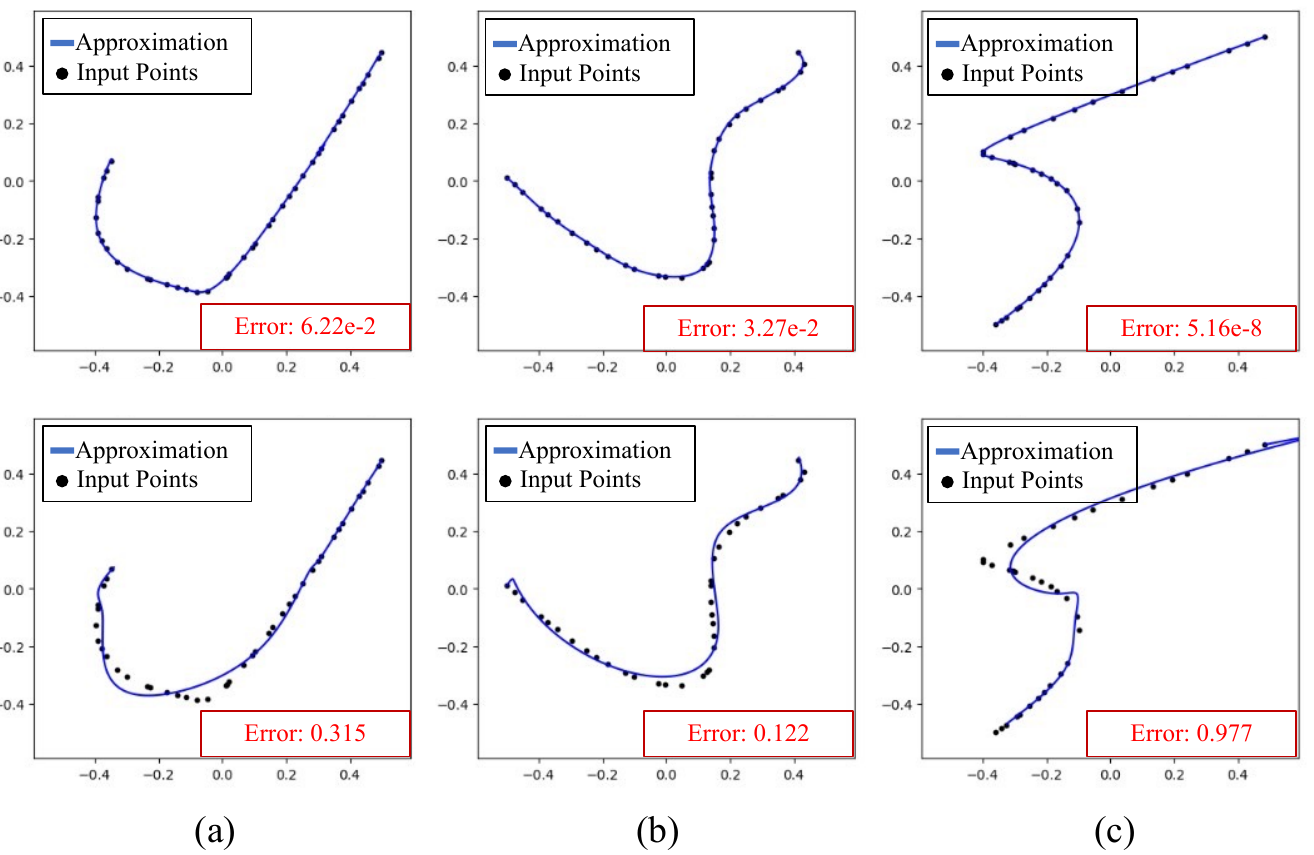}
    \caption{Influence of parameters, knots, and their alignment on approximation accuracy: (a) results from two different parameters; (b) results from two different knots; and (c) results of matched (top) and unmatched (bottom) cases.}
\label{fig:influence_of_params_and_knots}
\end{figure}

The above optimization problem is a least square problem, and its solution can be obtained by solving the following linear system:
\begin{equation}\label{eq:solution}
    (N^TN)P=N^TD
\end{equation}
where $P = [P_0,\dots,P_n]^T$, and
\begin{equation*}
    N=
    \begin{bmatrix}
        N_{0,p}(u_0) & \dots & N_{n,p}(u_0)\\
        \vdots & \ddots & \vdots\\
        N_{0,p}(u_m) & \dots & N_{n,p}(u_m)
    \end{bmatrix}
\end{equation*}
It is obvious that the coefficient matrix $N$ is solely determined by three factors:
\begin{enumerate}
    \item \textbf{Parametrization}. Each parameter $u_j$ indicates the position where to evaluate the B-spline basis function $N_{i,p}(u_j)$ and then determines all entries in the $j-$row of $N$. 
    Fig.~\ref{fig:influence_of_params_and_knots}a shows that parametrization can significantly affect the final result. 
        
    \item \textbf{Knot placement}. The knot vector $T$ determines each B-spline basis function $N_{i,p}(u)$, which in turn determines all entries in the $i-$column of $N$. 
    Fig.~\ref{fig:influence_of_params_and_knots}b shows that the choice of knot positions can affect the approximation accuracy a lot. 
        
    \item \textbf{Parameter-knot alignment}. Each entry of $N$ is collectively determined by parameters and knots. Leaning towards either side will not give good results.
    Fig.~\ref{fig:influence_of_params_and_knots}c shows that, if these two are not aligned well, a high approximation error occurs.
\end{enumerate}

Besides the above three factors, it is also needed to determine the number of knots, adaptively. When input data points do not assume any ordering a priori, the parameters' order also needs to be determined automatically. This does not mean sorting the parameter values, which is simple, but organizing the output parameters in a way consistent with the input data points.

\section{Methods}
\label{sec:methods}
\subsection{Network architecture}
\label{sec:architecture}
\begin{figure*}[t]
    \centering
    \includegraphics[width=0.9\textwidth]{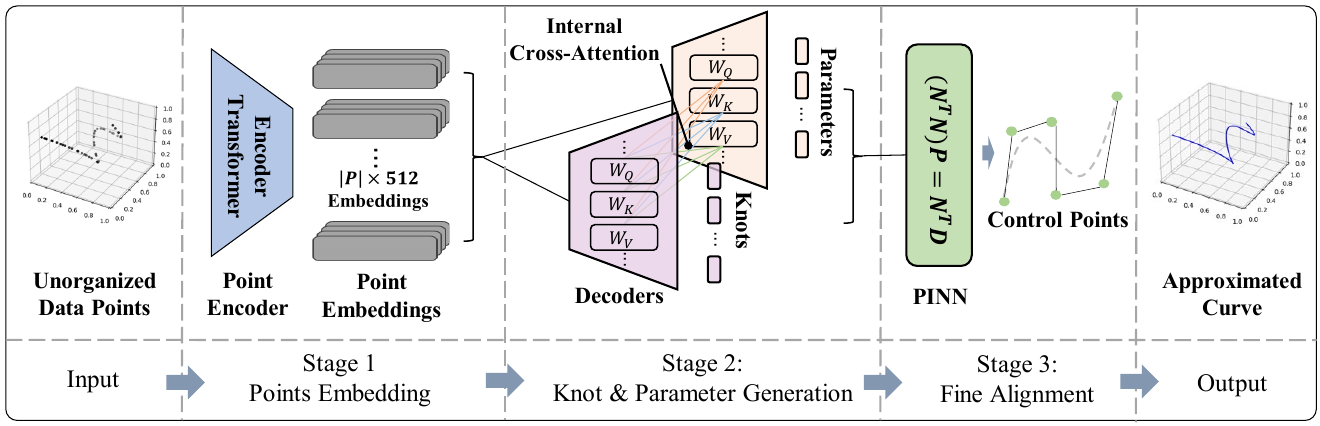}
    \caption{The overall architecture of SplineGen.} 
    \label{fig:network-architecture}
\end{figure*}

Rather than directly inputting raw point coordinates into the network (as in existing methods~\cite{laube2018deep,scholz2021parameterization}), SplineGen first seeks a neighborhood-aware high-dimensional embedding for each data point, see the left part of Fig.~\ref{fig:network-architecture}. This is done with the help of the attention mechanism in transformer~\cite{vaswani2017attention}---it allows every data point to attend to all data points in the input sequence. (This is the primary reason this work chooses transformer as the backbone.)  SplineGen combines it with an additional masking mechanism to force each data point to attend more to its neighborhood regardless of how points are organized. Specifically, we use masking to simulate point embedding corruption and ask the remaining point embeddings to yield the same curve as before. As such, each point embedding carries its neighborhood's information. To facilitate the alignment of parameters and knots, we force shared point embeddings between parameter generation and knot generation through the use of a shared autoencoder model. By doing so, we hope to capture the interplay between parameterization and knot placement in the high-dimensional embeddings. 

Having point embeddings in place, two decoders are used to decode them into knots and parameters, respectively. See the middle part of Fig.~\ref{fig:network-architecture}. The knot decoder autoregressively predicts the next knot based on already generated knots until it thinks enough of them are generated, indicated by an ``EOS" token. This essentially determines the number of knots and their placement simultaneously and automatically. Parameters are generated in a similar way, but with a self-organization module included in its decoder to ensure that the number of generated parameters and their order are in line with the input data points. To align these two generative models, a new module called internal cross-attention is added to the two decoders so that their attention modules are directly bundled together during knot and parameter generation. Lastly, the two decoders are further equipped with a masking mechanism to simulate data point removal/addition. It enforces a robust generation of knots and parameters (and therefore a robust approximation of the same curve) when some data points are missing or added.

SplineGen's last module is a physics-informed neural network (PINN) layer appended to the above two decoders. PINN is a machine learning technique allowing governing equations (e.g., Eqs.~\eqref{eq:optimizatioin} and~\eqref{eq:solution}) to be respected during training~\cite{prasad2022nurbs}. In this regard, the added PINN layer can directly use Eq.~\eqref{eq:solution} to guide the alignment of knot and parameter generation, capturing more subtle interplay between parameterization and knot placement than those obtained with shared embeddings and internal cross-attention. This module can be viewed as the final fine-tuning of the generative model.

\subsection{Learning point embeddings}
\label{sec:point-embedding}
\begin{figure}[t]
    \centering
    \includegraphics[width=0.48\textwidth]{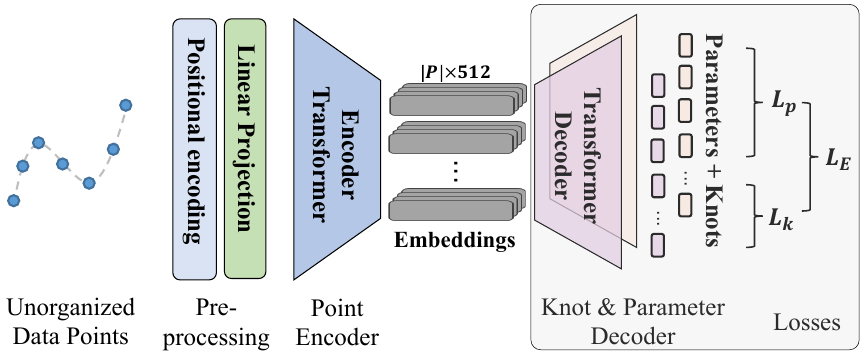}
    \caption{The network for learning point embeddings. The right grayed part is only used in the training phrase, not in the inference phrase.}
    \label{fig:point-embedding}
\end{figure}

The shared autoencoder model has three components, an encoder and two decoders, as shown in Fig.~\ref{fig:point-embedding}. The encoder takes as input the data points $P = \{p_i \in \mathbb{R}^3\}$ with their coordinates as 3-channel features for each point. These data points are not passed to the encoder directly but go through an intermediate step consisting of a positional encoding (on coordinates)  function $\Gamma: \mathbb{R}^3 \rightarrow \mathbb{R}^{d_{PE}}$~\cite{wang2021nerf} and a linear projection, as follows: 
\begin{equation*}
    h_P = \{p_i \rightarrow W_{P} \times \Gamma(p_i)\},
\end{equation*}
where $W_{P} \in \mathbb{R}^{d_{emb}\times {d_{PE}}}$ is a learnable matrix, with $d_{emb}=512$ and $d_{PE}=12$ for all cases in this work.

The above preliminary embeddings $h_P$ are then passed to a transformer encoder \textbf{E}, followed by a linear layer, to obtain the final point embeddings $E_P$:
\begin{equation*}
    E_P=\textbf{E}(h_P;\theta_{\textbf{E}})
\end{equation*}
where $\theta_{E}$ denote trainable parameters of the transformer encoder. Note that our transformer is not equipped with a position encoding, since input points are unorganized and the point embeddings should be permutation invariant. 

To train $E$ and obtain embeddings $E_P$ meaningful for parameterization and knot placement, an encoder-decoder scheme is used, with two transformer decoders: one for recovering knots from $E_P$, and the other for recovering parameters from $E_P$. Fig.~\ref{fig:point-embedding} shows the network. The specific procedures of training $E$ follow the autoregressive scheme~\cite{vaswani2017attention}. Specifically, assume that the two decoders have already generated $i$ knots and parameters, denoted by $\{\hat{k_0}, \dots, \hat{k_i}\}$ and $\{\hat{p_0}, \dots, \hat{p_i}\}$, they will predict the next knot and parameter from the same point embeddings $E_P'$ using:
\begin{equation}
    \begin{aligned}
        k_{i+1}&=\bar{\textbf{D}}_{K}(\{\hat{k}_0,\hat{k}_1,\dots,\hat{k}_i\},\textbf{E}_P';\theta_{\bar{\textbf{D}}_{K}},\theta_{\textbf{E}})\\
        p_{i+1}&=\bar{\textbf{D}}_{P}(\{\hat{p}_0,\hat{p}_1,\dots,\hat{p}_i\},\textbf{E}_P';\theta_{\bar{\textbf{D}}_{P}},\theta_{\textbf{E}})
    \end{aligned}
\end{equation}
where $\theta$ is the network parameters of its argument, \verb|<SOS>| is the starting token, $k_{i+1}$ and $p_{i+1}$ is the next predicted knot and parameter, respectively.

Instead of directly passing $E_{P}$ to the decoder, an additional masking mechanism is used to force each data point to attend more to its neighborhood regardless of how points are organized. Specifically, we randomly mask some embeddings to simulate point embedding corruption and ask the remaining point embeddings $E_{P}'$ to yield the same results as before (except for the masked embeddings' corresponding parameters). As such, each point embedding carries its neighborhood's information. 

The whole encoder-decoder pipeline is trained using the weighted sum of two losses, $L_{K}$ and $L_{P}$, which are the mean squared losses between ground truth($\hat{K}$ and $\hat{P}$) and predicted sequence($\bar{K}$ and $\bar{P}$), respectively.

\subsection{Generating knots and parameters}
\label{sec:generation}
\begin{figure*}[htbp]
    \centering
    \includegraphics[width=0.95\textwidth]{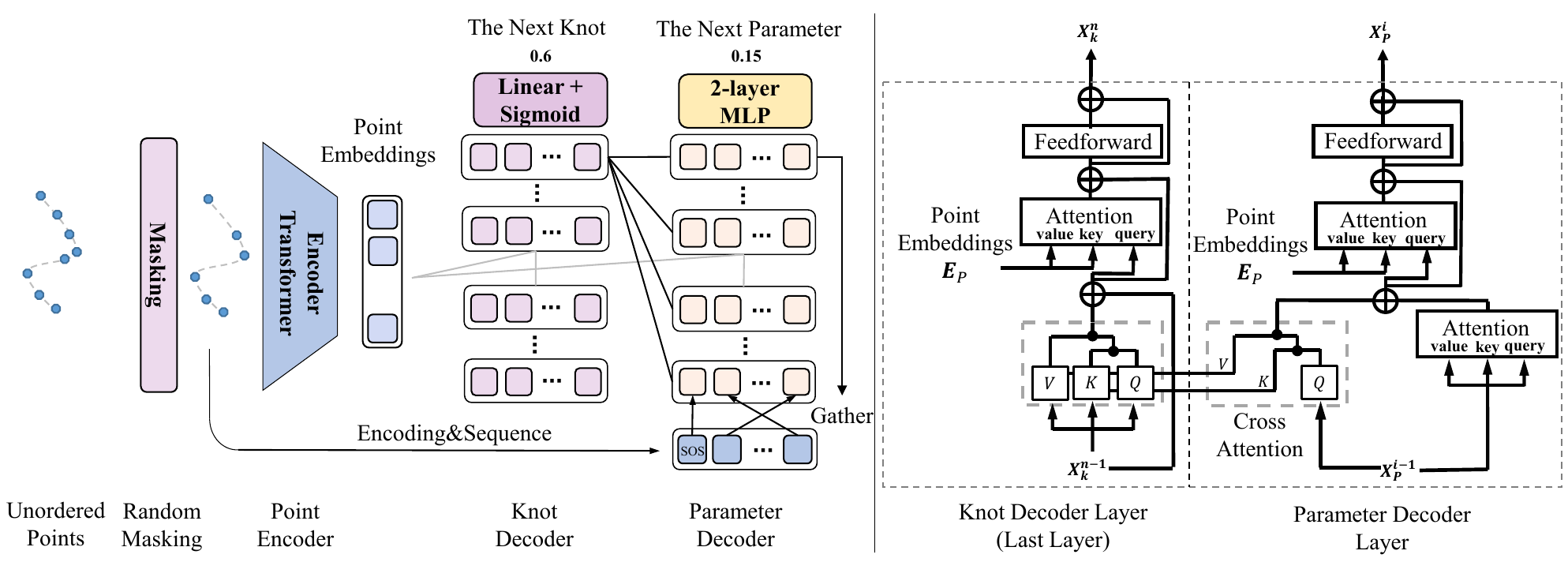}
    \caption{The network for generating knots and parameters. Left: The pipeline of knots and decoder generation. After random masking, the input points are sent to a transformer encoder, yielding the point embeddings. Both the knot decoder and parameter decoders take the embeddings for generation. Besides, point embeddings are gathered as input of the parameter decoder. For each layer of the parameter decoder, internal cross-attention is applied. Right: The details of the internal cross attention. The key and value in the self-attention module at the last layer of the knot decoder are used for the internal cross-attention with query vectors converted from $X_P^i$. Here, $X_k^{n}$ denotes the output of the last layer of the knot decoder, and $X_{P}^i$ denotes the output of $i^{th}$ layer of parameter decoder.}
    \label{fig:decoder}
\end{figure*}

From point embeddings, SplineGen decodes them into knots and parameters, again in an autoregressive way. Two problems need particular attention: how to ensure that the number of generated parameters and their ordering are in line with the input data points; and how to align parameterization and knot placement. This work solves these problems with several extensions to the transformer decoder. In a nutshell, the proposed decoder fuses two sub-decoders through a module called internal cross-attention (Fig.~\ref{fig:decoder}); one of the sub-decoders is a vanilla transformer decoder, and the other is a modified version with an additional self-organization module. 

\textbf{Knot sub-decoder.} SplineGen uses the same transformer decoder as the one used in Sec.~\ref{sec:point-embedding}. It generates knots autoregressively, starting from the \verb|<SOS>| token, predicting the next knot, and stopping at the \verb|<EOS>| token. The number of knots is determined by the network automatically.

To facilitate the training of this sub-decoder, we slightly modify the representation of knots. They are first sorted in non-decreasing order, then prefixed with the start token \verb|<SOS>| and suffixed with the stopping token \verb|<EOS>|. After inserting these special tokens, the knot sequence $K^{seq} = \{\verb|<SOS>|,k_0, k_1, \dots, k_{|K|-1},\verb|<EOS>|\}$ is of length $|K_{seq}| = |K| + 2$. Rather than using a real value for the knots, we add another two dimensions to indicate the probability of being \verb|<SOS>| and \verb|<EOS>| token, respectively.

Given all the point embeddings $E_P$ of given data points and the previously output knots $K^{seq}_{i-1}$ at step $i$, the goal is to model the next output knot token, denoted by $k_i$. The input to the transformer decoder is derived from each of the tokens in $K^{seq}_{i-1}$ by adding positional encoding followed by a linear project, and then the transformer decoder $\textbf{D}_{K}$ followed by another linear projection takes this input to predicts the next token. The model is trained to minimize the mean squared loss between predicted knots and the ground truth.

\textbf{Parameters sub-decoder.} For the parameter sub-decoder, the primary task is to make the output parameters' ordering in line with the input data points.
To do so, a parameter-point associative decoder is designed out of the transformer decoder used in Sec.~\ref{sec:point-embedding}. It adds an index prediction module so that the decoder can output not only the parameter values but also the positional indices of their corresponding points in the input. Specifically, the decoder takes the learned point embeddings $E_P$ as input into its cross-attention part and autoregressively outputs parameter values and indices. Let the decoder be denoted by $\textbf{D}_T(E_P;\theta_{\textbf{D}_T})$ and its output embeddings be $E_{T}$, we let each autoregressively generated embedding $E^i_{T}$ to learn a probability distribution $p_i$ over the indices of the point encodings $P'$ so that the one with the largest probability will correspond to the generated parameter in this iteration. 

A common way to define $p_i$ is expressing it in terms of a dot product operation between $E^i_{T}$ and encoded points $P'$, followed by a normalization operation with softmax, as follows~\cite{vinyals2015pointer}:
\begin{equation}\label{eq:index-probability}
     p_i = softmax((W_{Q}E^i_T)^{T} \cdot W_K P')
\end{equation}
where $W_Q \in \mathbb{R}^{d_{emb} \times d_{emb}}$ and $W_K \in \mathbb{R}^{d_{emb} \times d_{emb}}$ are learnable matrices. The encoded points $P'$ are simply latent codes of the input data points, and they are generated by a transformer encoder. The purpose of using $P'$ instead of the input data points is to add more trainable parameters.

Combining the trainable decoder $\textbf{D}_T$ and Eq.~\eqref{eq:index-probability} , the index prediction module essentially learns the following distribution:
\begin{equation}
p(I_t|E_P;\theta_{\textbf{D}_T},W_{Q},W_K)=\prod_{j=1}^{t}p(i_{j}|i_0,\dots,i_{j-1},E_P;\theta_{\textbf{D}_T},W_{Q},W_K),
\end{equation}
where $I_{t} = \{\verb|<SOS>|,i_0,i_1,\dots,i_{t}\}$ denotes the already predicted indices until step $t$ and $\theta_{\textbf{D}_T}$ is the decoder's parameters.

To avoid some indices being repeatedly predicted, we mask the corresponding entry in the dot product operation every time a new index is reached. This also guarantees that all elements in $P'$ will be visited because the prediction step is executed exactly $|E|$ times. 

To generate the intended parameter values, each embedding after the translation is passed to a 2-layer MLP to decode it into a parameter value. Combined with the previously generated index, we can obtain a parameter value, and meanwhile know its position in the final parameter array altogether. It should be noted that, to ensure the decoder generates as many parameters as the input data points, the generation process will be interrupted once enough parameters are output. This is different from the generation process for knots.

\textbf{Internal cross-attention.} Cross-attention has proven effective in model associativity between multimodal data~\cite{chen2021crossvit}. This inspires us to add an attention-based module to fuse the decoders described above and then align the knot generation and parameter generation processes. Different from most existing cross-attention models, which operate on an attention module's outputs (e.g.,~\cite{guo2022complexgen}), the internal cross-attention module here directly operates on its internal key/value/query matrices. This internal way of working can model the interplay between knots and parameters more conveniently.

The design of the internal cross-attention module is illustrated in Fig.~\ref{fig:decoder}. It activates only after the knots have all been generated. 
Upon reaching the final generation step of the knot sub-decoder ( with a token length denoted as $N_{k}$), both the key vectors $K\in \mathbb{R}^{N_{k} \times d_{attn}}$ and the value vectors $V\in \mathbb{R}^{N_{k} \times d_{attn}}$ have been generated in the self-attention module. These vectors are subsequently utilized in the upcoming internal cross-attention step. Consider the parameter generation at the time step $t$. At the $i^{th}$ level of the parameter sub-decoder, the output of $i-1^{th}$ decoder layer $X^{i-1}_P$ is transformed into query vectors $Q \in \mathbb{R}^{(t+1) \times d_{attn}}$ via a multiplication operation with a learnable matrix $W_{Q} \in \mathbb{R}^{d_{emb} \times d_{attn}}$. Employing these prepared keys, values, and queries within the cross-attention framework, an output vector is computed. This output is then combined with the preceding $X_P^{i-1}$ to yield a refined representation ${X^{i-1}_P}'$. The network module that follows uses this refined representation as its input instead of the original $X_P^{i-1}$.

\textbf{Masking.} Lastly, the two decoders are further equipped with a masking mechanism to simulate data point removal/addition. Different from the masking mechanism used in Sec.~\ref{sec:point-embedding}, here we randomly remove a small proportion of points from the input data while keeping the generated knots and parameters for the remaining data points unchanged. This method enforces a robust generation of knots and parameters when some data points are missing or added, thereby giving a robust approximation of the same curve.

\subsection{PINN-based fine alignment}
\label{sec:pinn}
To directly use Eq.~\eqref{eq:solution} to guide the alignment of knot and parameter generation, a PINN layer is further added to SplineGen when training it. This layer accepts the outputs from both the knot and parameter sub-decoders and computes a series of control points. The layer first calculates B-spline basis functions with an adapted version from the NURBS-Diff method presented in~\cite{prasad2022nurbs}, then solves Eq.~\eqref{eq:solution}, resulting in a set of control points $P$, which define the intended curve $C$. Utilizing the generated curve $C$, the parameters $u$, and the corresponding 3D points $D$, we evaluate the approximation error through the following loss function:
\begin{equation}
    \label{eq:loss}
    L = \max_{i=1}^{N} \| C(u_i) - D_i \|_2,
\end{equation}
where the maximum loss is employed instead of the mean loss to ensure that all predicted points consistently stay close to their target positions.

It should be noted that training a network with PINN layers is not easy. The derivatives of the PINN loss w.r.t the PINN input $u$ (i.e., parameters and knots in our case) need to be explicitly given to the network. The NURBS-Diff method~\cite{prasad2022nurbs} has provided a detailed derivation of those derivatives. We omit the details in this work.

\section{Results}
\label{sec:results}
\subsection{Setup}
\textbf{Dataset.} A dataset of 500,000 B-spline curves with their control points, knot vectors, and sampled points has been compiled. Self-intersecting curves were eliminated using a specialized detection program. For data processing, we normalize control points to $[0, 1]^3$ and utilize masked arrays for neural network training consistency, refer to Supplementary Material for more details. A test dataset of above 5,000 curves is generated in the same way. We will make these datasets publically available upon publication of this paper.

\textbf{Training.}
SplineGen has been implemented using PyTorch. We trained it on an NVIDIA RTX 3090 GPU and with the Adam solver, having a batch size 256 for 500 epochs and a fixed learning rate $lr$=$10^{-4}$. The dataset is split into 8:2 for training and validation.

\subsection{Evaluation Metrics}
Given the input data points $P$ and the reconstructed curve $C$ and predicted parameters $t$ for $P$, we evaluate the performance of SplineGen with the following three most commonly used metrics.

\textbf{Maxium of Euclidian distances}. We first evaluate the points in $C$ according to $t$ and compute the Euclidian distances of corresponding points. We use the max of these distances as our metrics, which is the same as Eq.~\eqref{eq:loss} to measure the maximum approximation distance:
  $$\max_{i=1}^{N} \| C(t_i) - D_i \|_2$$
  
\textbf{MSE of Euclidian distances}. In addition, we also evaluate the mean square of these distances:
  $$\frac{1}{N}\sum_{i=1}^{N} \| C(t_i) - D_i \|_2$$

\textbf{Hausdorff distances}. We also evaluate the Hausdorff distance between point set $P = \{C(t_i)\}_{i=0}^{N}$ and $Q = \{P_i\}_{i=0}^{N}$, which measures the greatest distance between any point in one set to its closest point in the other:
  $$\max\left\{\max_{p \in P}\min_{q \in Q} \| p - q \|_2,\max_{q \in Q}\min_{p \in P} \| p - q \|_2\right\}$$

\begin{figure*}[t]
    \centering
    \includegraphics[width=0.9\textwidth]{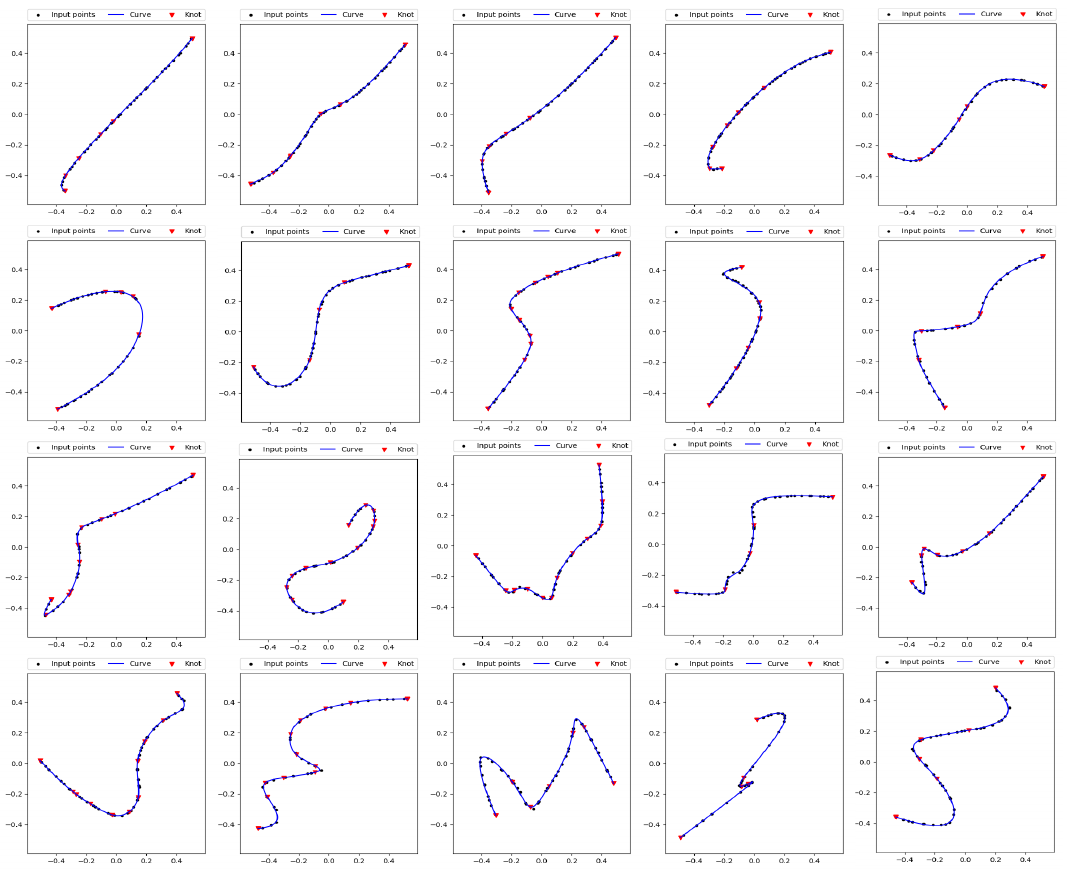}
    \caption{Results of 2D parametrization, knot placement, and B-spline approximation.}
    \label{fig:cases}
\end{figure*} 

\begin{figure*}[t]
    \centering
    \includegraphics[width=.9\textwidth]{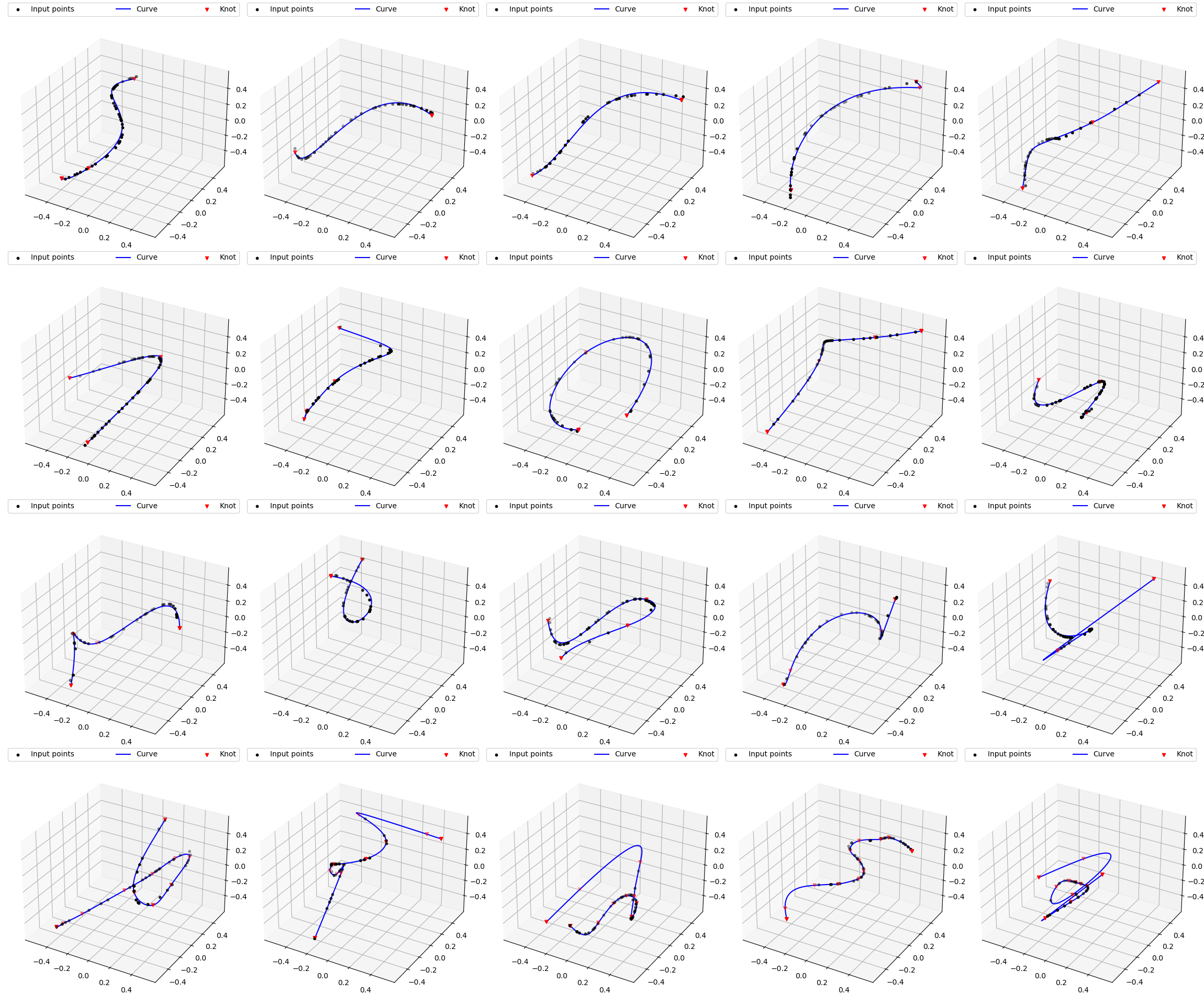}
    \caption{Results of 3D parametrization, knot placement, and B-spline approximation.}
    \label{fig:cases3d}
\end{figure*} 

\begin{figure*}[t]
    \centering
    \includegraphics[width=0.9\textwidth]{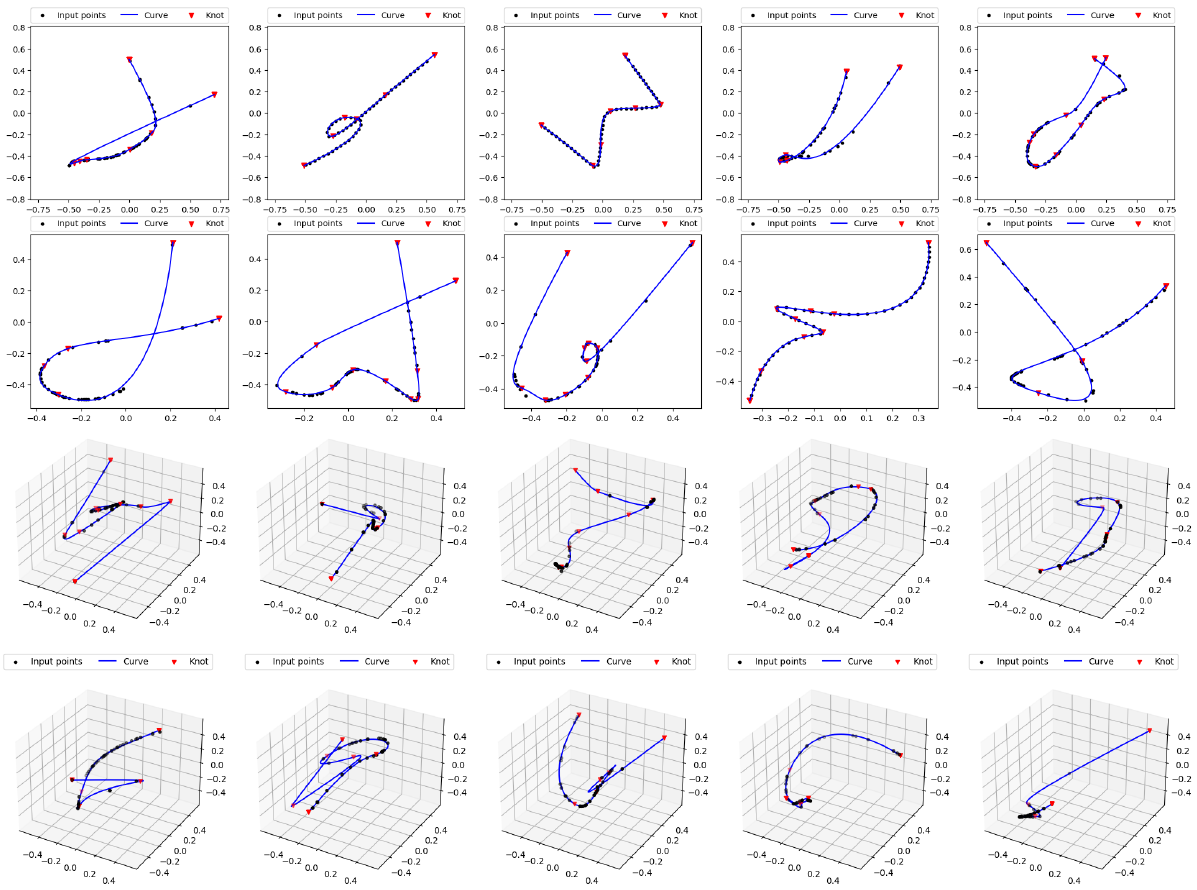}
    \caption{Results of approximating complex input data points. These shapes are rarely seen in practice, and they are only used to show the capability of SplineGen.}
    \label{fig:complex_case}
\end{figure*}

\subsection{Examples}
Some sampling results of using SplineGen to approximate data points are given in Figs.~\ref{fig:cases} and~\ref{fig:cases3d}, with increasing shape complexity from top to bottom. To further show the capability of SplineGen, Fig.~\ref{fig:complex_case} gives some approximation examples of very complex shapes. As can be seen from the figure, although these shapes are not practical and have unreasonably complex shapes, our SplineGen can still generate quality results.
  
\subsection{Ablation Studies}
An ablation study has been conducted on SplineGen's major modules, including the internal cross-attention module, the shared encoder module, and the masking module. Its setup is as follows:
\begin{enumerate}[label=(\alph*)]
    \item \label{ournet}\textbf{All:} The proposed SplineGen network, with all modules included.
    \item \textbf{w.o. Cross attention:} The internal cross-attention module is removed when training the parameter decoder, which means the parameter decoder is not informed with the knots information.
    \item \textbf{w.o. Shared encoder:} The shared point encoder module is replaced with a simple linear layer to encode the points.
    \item \textbf{w.o. Masking:} The masking module is removed when training the decoders.
\end{enumerate}

The ablation study results are shown in Table~\ref{tab:ablation}.
They show that SplineGen, only when all modules are included, can give the best performance. This confirms the effectiveness of the pipeline outlined in Fig.~\ref{fig:network-architecture}. Also, as each optionally removed module is related directly to the functions of parameter-knot alignment and robustness, the ablation study results validate SplineGen's features claimed in Sec.~\ref{sec:introduction}. Note that the features of size-independent and point permutation invariant are automatically ensured by the generative nature of SplineGen.

\begin{table}[t]
\caption{Stats of ablation studies.}
\centering
\setlength\extrarowheight{2pt}
\begin{tabularx}{0.5\textwidth}{
    >{\centering\arraybackslash}>{\hsize=1\hsize\linewidth=\hsize}X
    >{\centering\arraybackslash}>{\hsize=.5\hsize\linewidth=\hsize}X
    >{\centering\arraybackslash}>{\hsize=.5\hsize\linewidth=\hsize}X
    >{\centering\arraybackslash}>{\hsize=.5\hsize\linewidth=\hsize}X
    >{\centering\arraybackslash}>{\hsize=.5\hsize\linewidth=\hsize}X
    >{\centering\arraybackslash}>{\hsize=.5\hsize\linewidth=\hsize}X
   }
    \Xhline{1pt}
    Model &\makecell[c]{Param loss} &\makecell[c]{Ordering loss}\\
    \Xhline{0.5pt}
    All & 8.14e-3&3.18e-2\\
    \textbf{w.o.} Cross- attention&1.54e-2&3.38e-2\\
    \textbf{w.o.} Shared encoder& 9.90e-3 &3.58e-2\\    
    \textbf{w.o.} Masking& 8.68e-3 &3.83e-2\\
    \Xhline{1pt}
\end{tabularx}
\label{tab:ablation}
\end{table}


\begin{table*}[t]
\caption{Quantitative comparisons with existing methods, both learning-based methods and classical ones.}
\centering
\setlength\extrarowheight{2pt}
\begin{tabularx}{1\textwidth}{
    >{\centering\arraybackslash}>{\hsize=1\hsize\linewidth=\hsize}X
    >{\centering\arraybackslash}>{\hsize=.5\hsize\linewidth=\hsize}X
    >{\centering\arraybackslash}>{\hsize=.5\hsize\linewidth=\hsize}X
    >{\centering\arraybackslash}>{\hsize=.5\hsize\linewidth=\hsize}X
    >{\centering\arraybackslash}>{\hsize=.5\hsize\linewidth=\hsize}X
    >{\centering\arraybackslash}>{\hsize=.5\hsize\linewidth=\hsize}X
   }
    \Xhline{1pt}
    Model &\makecell[c]{Max Error} &\makecell[c]{MSE} &\makecell[c]{Hausdoff}\\
    \Xhline{0.5pt}
    SplineGen &\textbf{8.66e-3}&\textbf{2.81e-3}&\textbf{8.51e-3}\\
    J\"uttler~\cite{scholz2021parameterization} &7.71e-2&2.48e-2&5.54e-2\\
    PARNET~\cite{laube2018deep} &1.01e-1&2.24e-2&5.84e-2\\
    Centripetal+DOM &1.55e-1&8.04e-3&4.50e-2\\
    Chord+DOM &1.65e-1&1.01e-2&4.48e-2\\
    Centripetal+Uniform &6.01e-1&3.91e-2&1.44e-1\\
    Chord+Uniform &6.04e-1&4.11e-2&1.47e-1\\
    Centripetal+KTP &1.61e-1&1.21e-2&5.07e-2\\
    Chord+KTP &1.70e-1&1.04e-2&4.50e-2\\
    Centripetal+NKTP &1.57e-1&1.08e-2&4.71e-2\\
    Chord+NKTP &1.67e-1&9.06e-3&4.66e-2\\
    \Xhline{1pt}
\end{tabularx}
\label{tab:comparison2}
\end{table*}

\begin{table*}[th!]
    \caption{Sampling comparisons with existing methods on the curves of Fig.~\ref{fig:curves}.}
    \centering
    \setlength\extrarowheight{2pt}
    \begin{tabularx}{1\textwidth}{
        >{\centering\arraybackslash}>{\hsize=2.3\hsize\linewidth=\hsize}X
        >{\centering\arraybackslash}>{\hsize=0.8\hsize\linewidth=\hsize}X
        >{\centering\arraybackslash}>{\hsize=0.8\hsize\linewidth=\hsize}X
        >{\centering\arraybackslash}>{\hsize=0.8\hsize\linewidth=\hsize}X
        >{\centering\arraybackslash}>{\hsize=0.8\hsize\linewidth=\hsize}X
        >{\centering\arraybackslash}>{\hsize=0.8\hsize\linewidth=\hsize}X
        >{\centering\arraybackslash}>{\hsize=0.8\hsize\linewidth=\hsize}X
        >{\centering\arraybackslash}>{\hsize=0.8\hsize\linewidth=\hsize}X
        >{\centering\arraybackslash}>{\hsize=0.8\hsize\linewidth=\hsize}X
        >{\centering\arraybackslash}>{\hsize=0.8\hsize\linewidth=\hsize}X
        >{\centering\arraybackslash}>{\hsize=0.8\hsize\linewidth=\hsize}X
        >{\centering\arraybackslash}>{\hsize=0.8\hsize\linewidth=\hsize}X
        >{\centering\arraybackslash}>{\hsize=0.8\hsize\linewidth=\hsize}X       
        }
        \Xhline{1pt}
        Model & \multicolumn{3}{c}{Curve 1} & \multicolumn{3}{c}{Curve 2} & \multicolumn{3}{c}{Curve 3} & \multicolumn{3}{c}{Curve 4} \\
        \cline{2-4} \cline{5-7} \cline{8-10} \cline{11-13}

        & Max & Mse & Hsdff & Max & Mse & Hsdff & Max & Mse & Hsdff & Max & Mse & Hsdff\\
        \Xhline{1pt}
        
SplineGen & {\small \textbf{1.2e-2}}&{\small \textbf{2.7e-3}}&{\small \textbf{1.2e-2}}&{\small \textbf{1.1e-2}}&{\small \textbf{4.9e-3}}&{\small \textbf{1.1e-2}}&{\small \textbf{9.1e-3}}&{\small 3.3e-3}&{\small \textbf{7.7e-3}}&{\small \textbf{1.0e-2}}&{\small \textbf{4.7e-3}}&{\small \textbf{1.0e-2}} \\

J\"uttler~\cite{scholz2021parameterization} & {\small 4.7e-1}&{\small 2.9e-1}&{\small 8.4e-2}&{\small 4.2e-1}&{\small 2.2e-1}&{\small 8.4e-2}&{\small 3.8e-1}&{\small 1.2e-1}&{\small 8.0e-2}&{\small 3.6e-1}&{\small 2.0e-1}&{\small 6.9e-2} \\
PARNET~\cite{laube2018deep} & {\small 3.6e-2}&{\small 1.2e-2}&{\small 3.6e-2}&{\small 6.5e-2}&{\small 1.9e-2}&{\small 4.5e-2}&{\small 8.8e-2}&{\small 1.7e-2}&{\small 8.8e-2}&{\small 4.9e-1}&{\small 3.4e-2}&{\small 2.0e-1} \\
Centri + DOM & {\small 2.8e-2}&{\small 8.9e-3}&{\small 2.8e-2}&{\small 5.0e-2}&{\small 1.3e-2}&{\small 5.0e-2}&{\small 2.3e-2}&{\small 5.2e-3}&{\small 2.3e-2}&{\small 5.1e-1}&{\small 2.2e-2}&{\small 2.0e-1} \\
Chord + DOM & {\small 6.5e-1}&{\small 2.6e-2}&{\small 7.4e-2}&{\small 4.7e-2}&{\small 1.2e-2}&{\small 4.7e-2}&{\small 6.7e-1}&{\small 3.0e-2}&{\small 1.1e-1}&{\small 2.6e-2}&{\small 8.9e-3}&{\small 2.6e-2} \\
Centri + Uniform & {\small 3.3e-2}&{\small 7.9e-3}&{\small 3.3e-2}&{\small 3.1e-2}&{\small 8.8e-3}&{\small 3.1e-2}&{\small 9.3e-3}&{\small \textbf{3.2e-3}}&{\small 9.3e-3}&{\small 5.1e-1}&{\small 1.8e-2}&{\small 2.0e-1} \\
Chord + Uniform & {\small 6.5e-1}&{\small 2.6e-2}&{\small 7.4e-2}&{\small 3.3e-2}&{\small 8.2e-3}&{\small 3.3e-2}&{\small 6.7e-1}&{\small 2.7e-2}&{\small 1.1e-1}&{\small 1.9e-2}&{\small 5.1e-3}&{\small 1.9e-2} \\
Centri + KTP & {\small 3.6e-2}&{\small 1.3e-2}&{\small 3.6e-2}&{\small 5.2e-2}&{\small 1.5e-2}&{\small 5.2e-2}&{\small 2.8e-2}&{\small 6.7e-3}&{\small 2.8e-2}&{\small 5.1e-1}&{\small 2.2e-2}&{\small 2.0e-1} \\
Chord + KTP & {\small 6.5e-1}&{\small 3.1e-2}&{\small 7.7e-2}&{\small 5.4e-2}&{\small 1.4e-2}&{\small 5.4e-2}&{\small 6.7e-1}&{\small 2.7e-2}&{\small 1.0e-1}&{\small 3.3e-2}&{\small 1.0e-2}&{\small 3.3e-2} \\
Centri + NKTP & {\small 3.0e-2}&{\small 1.0e-2}&{\small 3.0e-2}&{\small 4.7e-2}&{\small 1.3e-2}&{\small 4.7e-2}&{\small 2.4e-2}&{\small 6.9e-3}&{\small 2.4e-2}&{\small 5.1e-1}&{\small 2.2e-2}&{\small 2.0e-1} \\
Chord + NKTP & {\small 6.5e-1}&{\small 2.8e-2}&{\small 7.4e-2}&{\small 4.5e-2}&{\small 1.2e-2}&{\small 4.5e-2}&{\small 6.7e-1}&{\small 2.7e-2}&{\small 1.1e-1}&{\small 2.3e-2}&{\small 9.6e-3}&{\small 2.3e-2} \\

\Xhline{1pt}
\end{tabularx}
\label{tab:comparisons}
\end{table*}

\subsection{Comparisons}
\label{sec:comparison}
Table~\ref{tab:qualitative comparison} shows a qualitative comparison between SplineGen with existing learning-based methods (i.e., J\"uttler~\cite{scholz2021parameterization} and PARNET~\cite{laube2018deep}). SplineGen is the only method that can generate parameters and knots with unordered and non-fixed-size inputs.
Table~\ref{tab:comparison2} presents a quantitative comparison of SplineGen with existing methods. Considering that all those existing methods require ordered data points as input, we have pre-sorted the input accordingly to make the comparison fair. Besides learning-based methods, we also compared with various classical parametrization and knot placement methods. Specifically, the chord-length and centripetal methods have been chosen for parametrization; NKTP~\cite{piegl2000surface}, KTP~\cite{PiegTill1996}, DOM~\cite{park2007b} and uniform knots~\cite{PiegTill1996} have been chosen for knot placement. All of their combinations have been analyzed, as shown in Table~\ref{tab:comparison2}.

\begin{figure*}[htbp]
    \centering
    \includegraphics[width=0.97\textwidth]{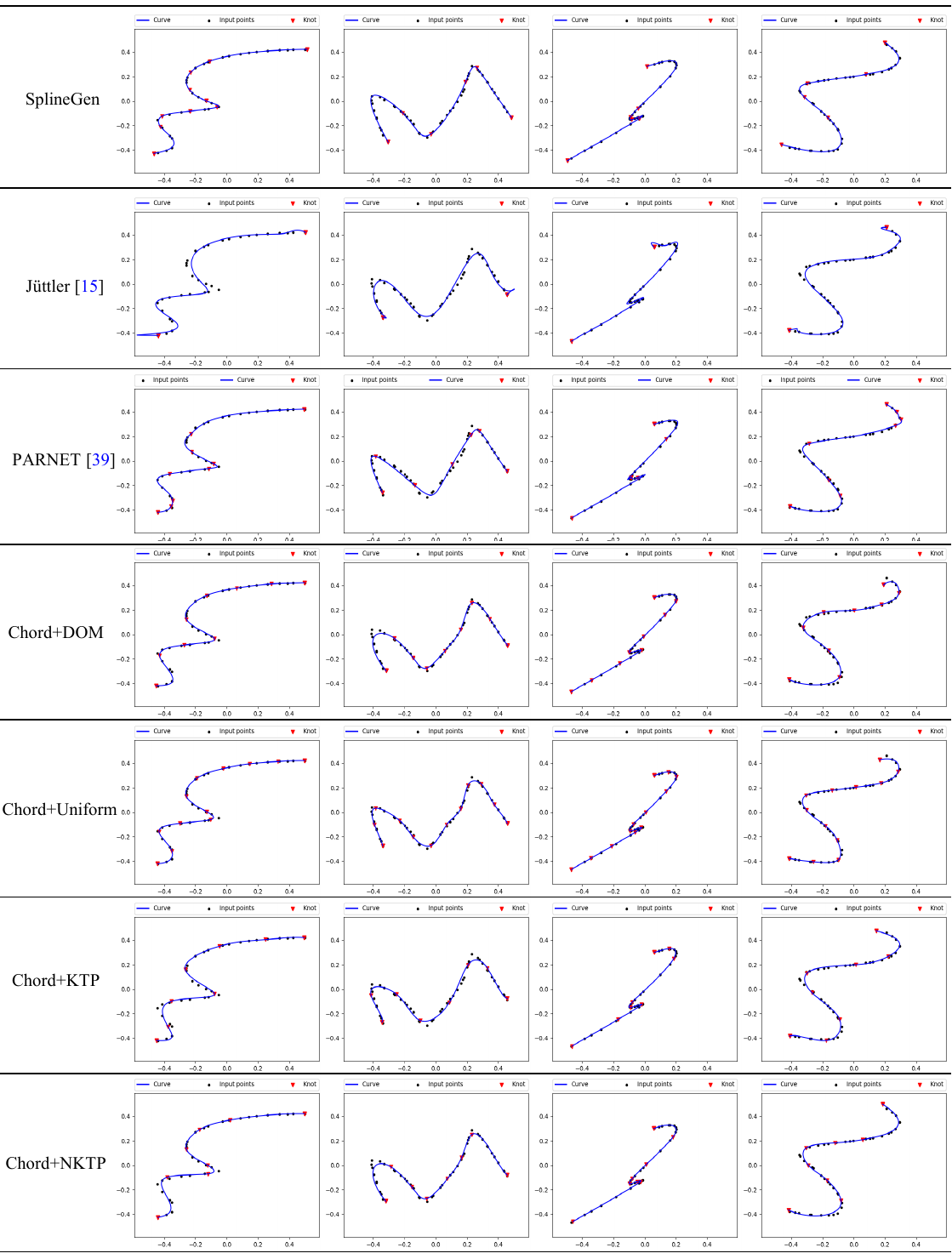}
    \caption{Comparisons of Spline with various existing methods on curves from Fig.~\ref{fig:cases}.}
    \label{fig:curves}
\end{figure*}

The results indicate that SplineGen outperforms other methods in terms of maximum error (Max), mean squared error (Mse), and Hausdorff error (Hsdff).
Table~\ref{tab:comparisons} further provides statistics of comparing SplineGen with existing methods on four curves chosen from the examples in Fig.~\ref{fig:cases}. Fig.~\ref{fig:curves} shows the resulting approximate B-spline curves from these methods.
SplineGen is seen to give competitive results in terms of both accuracy and generality. Only in the Mse error of Curve 3, SplineGen is less accurate than the method using the centripetal method for parametrization and the uniform method for knot placement. However, the result of SplineGen is very close to the best result, with only a 1e-4 difference.

\subsection{Discussion and limitations}
From the comparisons shown in Figs.~\ref{fig:cases}-\ref{fig:curves}, and Tables~\ref{tab:comparison2} and~\ref{tab:comparisons}, SplineGen can consistently give competitive approximation results, with lower error and better robustness. Large improvements have been confirmed for classic parametrization and knot placement methods (except for one case), see Fig.~\ref{fig:curves}. Even for the state-of-the-art learning-based method, the proposed method is still able to reduce the maximum error by a notable percentage, above 88\%, depending on the specific data points being considered.

Numerical issues in solving Eq.~\eqref{eq:solution} have also been observed. An in-depth analysis showed that the reason lies in the distribution of the parameters generated by SplineGen. It has been known that it is better to have parameters filled in every knot span, and empty knot spans will result in a poorly conditioned coefficient matrix~\cite{farin2002curves}. To completely solve this issue, a network that can evenly distribute parameters over knot spans is needed. Such a network remains unknown, and further development is required.

\section{Conclusion}
\label{sec:conclusion}
A generative model called SplineGen has been presented to solve the traditional parameterization and knot placement problems in B-spline approximation. The main features of this model include input data point size-independent, point permutation invariant, aligned knots and parameters, and robustness, cumulatively giving a highly accurate parameterization and knot placement method. These features are essentially attained by casting the parameterization and knot placement problems as a sequence-to-sequence translation problem, together with several extensions to the existing transformer network to account for the special needs of parameterization and knot placement. A notable improvement over existing methods has been demonstrated in various experiments.

It should, however, be noted that SplineGen only represents a first step towards generative spline modeling, and much more work remains to be done. Most notably, extending SplineGen to the problem of surface fitting of unorganized points is among the most important improvement directions. Our preliminary results show that the primary challenge in this extension is to handle scattered data points assuming no tensor product grids or rectangular topology (i.e., the underlying surfaces are trimmed splines). There is no obvious solution to this challenge, and this seems to be the main reason why there is no successful application of learning-based methods to surface fitting. 

For the current SplineGen, there is still room for improvement. During the implementation, it was found that a small error in the parameter-point associative module could yield a large error in the final results. This is mainly because SplineGen uses a global way to generate the associativity. This will have issues when the number of input data points is large. For such cases, a parameter-point associative network that can attention to local neighborhoods is preferred. Then those local associativity can be assembled to attain the overall associativity. Developing such a network module is practically beneficial. Another interesting improvement lies in controlling the distribution of generated parameters. It is known that good parameterization distributes parameters evenly in individual knot spans; otherwise, numerical issues would happen when solving Eq.~\eqref{eq:solution}. Incorporating this knowledge into SplineGen is among the research studies to be carried out in the future. 

\section*{Acknowledgements}
This work has been funded by NSF of China (No. 62102355), the ``Pioneer" and ``Leading Goose" R\&D Program of Zhejiang Province (No. 2024C01103), NSF of Zhejiang Province (No. LQ22F020012), and the Fundamental Research Funds for the Zhejiang Provincial Universities (No. 2023QZJH32).

\section*{References}
\bibliographystyle{elsarticle-num}
\bibliography{mybibfile}

\end{document}